\documentclass[sigconf,screen,nonacm]{acmart}
\usepackage{geometry,setspace,hyperref,graphicx,float,algorithm,algpseudocode, tabularx, amsmath,subcaption, tcolorbox}

\tcbset{
    sharp corners,
    colback = white,
    before skip = 0.2cm,    
    after skip = 0.5cm      
}   

\newtcolorbox{boxA}{
    fontupper = \bf,
    boxrule = 1.5pt,
    colframe = black 
}

\newboolean{showcomments}
\setboolean{showcomments}{true}         
\ifthenelse{\boolean{showcomments}}
  {\newcommand{\nb}[2]{
  \fbox{\bfseries\sffamily\scriptsize#1}
     {\sf\small$\blacktriangleright$\textit{\textcolor{red}{#2}}$\blacktriangleleft$}
   }
  }
  {\newcommand{\nb}[2]{}
   
  }

\author{Kuen Sum Cheung}
\affiliation{%
  \institution{King's College London}
  \city{London}
  \country{UK}
}
\email{kuen.cheung@kcl.ac.uk}

\author{Mayuri Kaul}
\affiliation{%
  \institution{King's College London}
  \city{London}
  \country{UK}
}
\email{mayuri.kaul@kcl.ac.uk}

\author{Gunel Jahangirova}
\affiliation{%
  \institution{King's College London}
  \city{London}
  \country{UK}
}
\email{gunel.jahangirova@kcl.ac.uk}

\author{Mohammad Reza Mousavi}
\affiliation{%
  \institution{King's College London}
  \city{London}
  \country{UK}
}
\email{mohammad.mousavi@kcl.ac.uk}

\author{Eric Zie}
\affiliation{%
  \institution{Charsfield Research \& Advisory}
  \city{London}
  \country{UK}
}
\email{eric.zie@cranda.digital}

\begin{document}

\title{Comparative Analysis of Carbon Footprint in Manual vs. LLM-Assisted Code     Development}

\begin{abstract}
Large Language Models (LLM) have significantly transformed various domains, including software development. These models assist programmers in generating code, potentially increasing productivity and efficiency. However, the environmental impact of utilising these AI models is substantial, given their high energy consumption during both training and inference stages. This research aims to compare the energy consumption of manual software development versus an LLM-assisted approach, using Codeforces as a simulation platform for software development. The goal is to quantify the environmental impact and propose strategies for minimising the carbon footprint of using LLM in software development. Our results show that the LLM-assisted code generation leads on average to 32.72 higher carbon footprint than the manual one. Moreover, there is a significant correlation between task complexity and the difference in the carbon footprint of the two approaches.  
\end{abstract}



\maketitle

\section{Introduction}
Large Language Models (LLM) are disrupting many sectors \cite{Makridakis2023} and providing novel ways of producing different types of content \cite{Deloitte2023}. The impact of LLM has expanded to software development and a myriad of approaches have emerged to assist and automate various software development tasks, such as design, implementation, and testing \cite{10.1007/978-3-031-46002-9_23}\cite{Hou2023}.   Since their inception, there has been a genuine concern about the environmental impact of LLM, including their carbon footprint \cite{faiz2024llmcarbon}. Several studies have raised ethical concerns and showed that LLM use a massive amount of energy. For instance, the end-to-end carbon footprint of GPT3 is estimated to be 554 tons of ${\mathrm{CO}}_2$ equivalent \cite{faiz2024llmcarbon}. This study aims to provide a quantitative analysis of the environmental impact of LLM in programming tasks, when used in an assistive mode and compare it to the traditional manual process of coding.  

To make this feasible, we focus on programming tasks for which there are large datasets of programming effort available from public platforms. We then run the same tasks in the LLM-based process and compare the efficiency, correctness, and end-to-end estimated energy consumption of the two approaches. Using these metrics, we 
 answer the following research questions.

\begin{enumerate}
    \item[\textbf{(RQ1)}] {Does LLM-based software development lead to less carbon emissions than manual software development?} \label{RQ1}
\item[\textbf{(RQ2)}]{Is there a correlation between the complexity of the requirements and the difference in carbon footprint between the LLM-based and manual approach? Here, by ‘complexity’ we mean the difficulty level assigned by Codeforces to a given problem and ‘requirement’ refers to the set of instructions that define the task in a Codeforces problem statement.}\label{RQ2}  
\end{enumerate}

The overall objectives of our study are: 1) to thoroughly analyse the sustainability of LLM-based software development in comparison to manual-based software development, and 2) to propose strategies for the best-practice use of LLM in software development.

The findings of this paper indicate that the end-to-end energy consumption of the LLM-based process is an order of magnitude larger than that of the manual process in all our sampled tasks. Moreover, the gap between the energy consumption of the two approaches increases significantly with the complexity of the requirements. Our results indicate the necessity of a broader study in order to find processes in which the LLM-based approaches may have a comparable or lower energy consumption to justify their replacement for the manual process. We provide the code we have used to calculate the metrics as well as our full experimental data as part of our replication package~\cite{replpackage}.

\section{Literature Review} \label{sec:review}
In this section, we review the literature on the carbon footprint of software development, with a focus on the emerging role of LLM. We provide an overview of the key relevant studies and identify any gaps in the current body of research. 

\noindent \textbf{Carbon Footprint of Software Development.} There is a substantial and growing  interest in assessing the carbon footprint of traditional software development; we refer to surveys \cite{Karita2019,Venters2023}, scoping- \cite{Swacha2022,Ibrahim2022,McGuire2023}, and  mapping studies \cite{bambazek_requirements_2023}. 
Recent work indicates a lack of awareness and evaluation methods as major difficulties in adopting sustainable practices in software development \cite{Karita2019,Konig2024,Wahler2024}.  We address these by providing a basic methodology to evaluate and compare LLM-based software development with manual development. Moreover, we raise awareness by demonstrating  significant, and to our knowledge hitherto unknown, differences in the carbon footprints of the two approaches.   

\noindent \textbf{Carbon Footprint of LLM-Assisted Software Development.}
 To date, there have been few studies addressing the carbon footprint of LLM in software development. 
Faiz et al.\  explore the substantial carbon footprint generated by LLM during training and inference \cite{faiz2024llmcarbon}. The end-to-end carbon footprint of GPT3 is estimated to be 554 tons of ${\mathrm{CO}}_2$ equivalent which indicates the high environmental cost of LLM-assisted programming.
Additionally, a recent line of work indicates that the code generated by the LLM has a larger carbon footprint than the human-written code \cite{Tina2024,Vartziotis2024}.   This line of work differs from our work in that it focuses on the sustainability metrics of LLM-generated code rather than evaluating the carbon footprint of the development effort. The work by Belchev \cite{belchev2025code} is one of the first works that investigates this direction with a focus on the LLM's energy consumption and efficiency in software development tasks such as code generation, bug fixing,
documentation, and testing. However, this work does not provide a comparison to the carbon footprint of the traditional, human-led development efforts.

\noindent \textbf{Identified Research Gaps.}
While there is increasing interest in studying both the carbon footprint of traditional software development and the environmental impact of LLM, there is a notable gap in research focused on the carbon footprint of  LLM-assisted software development and how it compares to the manual software development. The broader context of using LLM throughout the software development lifecycle remains underexplored. Our research aims to fill this gap.

\section{Programming Tasks Dataset}\label{sec:dataset_selection}
In this section, we describe the dataset selection process and  criteria for task inclusion.

\subsection{Choice of Programming Tasks Dataset}
To answer our research questions, we include programming  task datasets satisfying the following two essential criteria: 
\begin{enumerate} 

\item \textbf{Well-Defined and Diverse Tasks:} We required tasks that are clearly defined 
and cover a range of complexities. In particular, every task is accompanied by clearly defined test cases and explicit instructions, which provide concrete examples of expected functionality and performance benchmarks. Well-defined tasks ensure that the LLM focuses on solving the task without performance loss due to ambiguity. Additionally, incorporating tasks of varying complexity allows us to measure how the carbon footprint scales with the difficulty of the tasks, providing insights into the energy efficiency of the LLM across different types of challenges.

\item \textbf{Comprehensive Submission Data:} Beyond just the tasks, we needed a substantial amount of data, including detailed submission records from at least 1,000 participants who successfully solved each task to avoid bias, account for variability in human performance, and gain statistical confidence. This data should include crucial metrics such as code runtime, memory usage, and the time spent on each task. We would prefer a platform that provides an API to facilitate the automation of our process. 

\end{enumerate}

When choosing a dataset for programming tasks, multiple platforms were considered, including LeetCode \cite{leetcode2024}, HackerRank \cite{hackerrank2024}, and Codeforces \cite{codeforces2024}. Each of these platforms hosts competitive programming contests and provides a range of task complexities, making them potential candidates for our research. However, the final selection was narrowed down to Codeforces because none of the other platforms, e.g., LeetCode and HackerRank, provide the data needed for our study, particularly regarding the time spent on each task. This data is essential for estimating the task complexity and calculating the associated carbon footprint.
Codeforces, however, provides the data through an API facilitating the automation of our measurement and comparison process. 

Codeforces is one of the most popular platforms for practicing competitive programming, attracting both novice and experienced programmers from around the world. 
Below we provide some key concepts used in the remainder of our methodology:

\begin{itemize} 
\item \textbf{Contests}: Users participate in timed contests, which feature a suite of well-defined programming tasks of varying complexity. 
These contests are numbered as rounds;  in our experiment, we selected rounds 1983, 1984, and 1994 for analysis, which have sufficiently many tasks of varying complexity with sufficiently many participants   (12 tasks in total).

\item \textbf{Rating System}\label{rating}: After participating in a contest, a user's rating will increase or decrease based on their performance relative to other contestants. With a current range of ratings from 4009 to -53, this rating system confirms that both novice and experienced programmers actively use Codeforces.
\item \textbf{Editorials}: After a contest has ended, the organizers release code solutions and detailed editorials explaining how to solve each task. These editorials were used 
as a part of the prompt to the LLM. 
\end{itemize}

\subsection{Task Selection}

Within Codeforces not all tasks had the desired number of submissions or included the required data, such as code runtime or memory usage. We focused on tasks that met the following conditions: 

\begin{enumerate} 
\item Each included task should be solved by at least 1,000 participants to ensure diversity and statistical significance. In our study, we assume that the programming experience held by the participants is normalized across participants based on the variety of user ratings in the contest.
 
\item Each included task should provide code runtime and memory usage.
\end{enumerate}

\subsection{Justification for Python-Specific Focus}

Python was chosen as the primary language for this study for two main reasons. First, GPT-4 outputs Python by default when generating code, making it critical to assess LLM performance in the language it most frequently uses. Second, Python is the second most popular language used on Codeforces, with a wide range of participants consistently submitting solutions in Python during contests. This popularity ensures that we have access to a large dataset of Python submissions, providing the necessary metrics to evaluate both runtime performance and carbon footprint.

\section{LLM-Assisted Code Generation}\label{sec:llm_codegen}

\begin{figure*}[!t]
    \centering
    \includegraphics[width=0.8\linewidth]{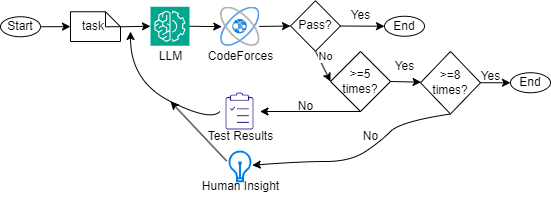}
    \caption{One Iteration of the LLM Experiment}
    \label{fig:experiment}
\end{figure*}

To assess the carbon footprint of the LLM-assisted approach, we need to simulate software development using LLMs. For this, we have designed the following process:

\begin{enumerate}
    \item We provide the problem to the LLM and ask the LLM to write the Python solution for it.
    \item We submit the code generated by the LLM to Codeforces, and record the number of tests passed. %
    \item If any of the test cases fail, we provide the LLM with the error trace of the failing test cases and ask it to fix the code.
    We repeat this step up to 4 times. This number is chosen as a practical balance: it gives the LLM several chances to correct its mistakes while keeping the overall process efficient, given that improvements tend to taper off after a few iterations.
    \item If the LLM still cannot generate code that passes all test cases after 5 queries, we assume that it is not capable of solving this task without human support. To provide such support, we pass the LLM the \textit{human insight} provided by Codeforces for this specific task. This insight is a short textual guide for the programmer on how to approach the task.    

    \item If the LLM still fails, we feed it the test result, keep the human insight in the prompt and ask the LLM to fix the code. We repeat this step for up to 3 times. Similarly, the limit of 3 iterations in this phase is set to offer the LLM enough opportunities to integrate the additional human guidance while maintaining a feasible and controlled experimental framework.
\end{enumerate}

For each task, we repeat the process starting from Step 1 three times and record the mean number for each metric. Our study uses 3 contests which have 12 valid tasks. Figure \ref{fig:experiment} demonstrates one repetition of the experiment for one task.

The LLM we choose in this study is GPT-4. We chose this model because we found sufficient data on training and query costs, which are relevant to our energy estimation. For most other popular (e.g., programming-focused) LLM, such data is not available.

\section{Carbon Footprint Estimation}
\label{sec:experiment-design}

In this section we introduce the key terms and formulas required for carbon footprint estimation and discuss the values that we used for various metrics. We then describe how we estimated the carbon footprint for manual and LLM-assisted code generation.

\subsection{Key Terminology, Formulas and Metric Values}

To provide details on carbon footprint estimation that we have performed as part of this study, it's essential to define key terms. The \textbf{carbon footprint (CF)} represents the total greenhouse gas emissions caused directly or indirectly by a system, typically measured in kilograms of $CO_2$ equivalent ($kgCO_2$$e$). In the context of software development, this footprint is tied to the total energy consumed (TTEC) during execution. The \textbf{consumed energy ($E$)} is the product of the power of a device ($p$) and the time ($t$) it runs and is typically measured in kilowatt-hours ($kWh$):

\begin{equation}
    E = p \times t
    \label{eq:energy_formula}
\end{equation}

\textbf{Carbon intensity (CI)} refers to the amount of carbon dioxide ($CO_2$) emissions produced per unit of energy consumed, typically measured in kilograms of $CO_2$ equivalent per kilowatt-hour ($kgCO_2$$e$/$kWh$). The carbon intensity of an energy source is a key factor in determining the environmental impact of energy consumption because different energy sources emit different amounts of $CO_2$ when generating electricity. For instance, fossil fuels have high carbon intensities because burning them releases a significant amount of carbon dioxide, while renewable energy sources such as wind, solar, or hydro have much lower carbon intensities, as they do not produce $CO_2$ when generating electricity. It should also be noted that the carbon intensity of energy sources is location-sensitive and can further be refined if information about the geographical location of energy sources is provided.

The carbon footprint (CF) of software can be calculated using the following formula:
\begin{equation}
    CF = E \times CI
    \label{eq:carbon_footprint}
\end{equation}
where $E$ is the consumed energy and $CI$ is the carbon intensity of that energy.

As the approximation for carbon intensity, in this study we use the data provided by the 2023 report from Nowtricity~\cite{nowtricity2024}. This report estimates emissions based on the life cycle $CO_2$ equivalent for each energy source. Nowtricity uses methods and data defined in UNECE~\cite{unece2022} and IPCC~\cite{ipcc2014} reports, including not just direct emissions but also those from infrastructure and the supply chain. Based on all this data, Nowtricity reports the average emissions of 217g $CO_2$ per $kWh$ which is the number we use in all of our calculations. 

As Formula~\ref{eq:energy_formula}  indicates, to compute energy consumption, we need to know the values for power ($p$). In our study, we assume that the code generation is performed using a standard laptop. We distinguish between the power the laptop consumes during regular use for coding activities ($p_{\text{laptop}}$) and the power it draws when running the code ($p_{\text{runtime}}$). As the value for $p_{\text{laptop}}$ we use the 4.075W, the average between the values of 3.59W (reported by Asus~\cite{asus2024}) and of 4.62W (reported by Dell~\cite{dell2024}).

We define $p_{\text{runtime}}$ as the combined power of the CPU ($p_{\text{cpu}}$) and RAM ($p_{\text{ram}}$), adjusted by the memory usage percentage ($u$):
\begin{equation}
    p_{\text{runtime}} = p_{\text{cpu}} + p_{\text{ram}} \times u
    \label{eq:runtime_power}
\end{equation}

In our study, we assume that a RAM of 16 GB is being used, and therefore estimate $u$ as the used RAM divided by 16.
We measure $p_{\text{cpu}}$ and $p_{\text{ram}}$ using the Python library CodeCarbon \cite{codecarbon} by running the code locally. However, it was observed that the CPU power usage remained consistent and was capped at the Thermal Design Power (TDP) of the processor. Similarly, runtime RAM power is also capped to a certain value.

\subsection{Carbon Footprint for Manual Approach}

In this subsection, our goal is to estimate the \textit{carbon footprint} of the manual code generation for each task. We calculate the \textit{total energy consumption} as the sum of three distinct parts: \textit{Coding Energy Consumption} (CEC), \textit{Debugging Energy Consumption} (DEC) and \textit{Testing Energy Consumption} (TEC).

\noindent \textbf{Coding Energy Consumption (CEC)}: 
This metric refers to the energy consumed by an average laptop during the coding process, excluding energy used for testing and debugging. We calculated it using the Formula~\ref{eq:energy_formula} and $p_{laptop}$ value for the power. 

To calculate the mean time spent (MTS) on coding for each task denoted by $t$, we used the relative submission time provided by Codeforces. The relative submission time is the timestamp of when a participant submitted a solution for a task, measured from the start of the contest. Since Codeforces only provides this relative submission time (rather than the actual time spent on each task), we had to take a number of measures explained below to ensure accuracy.

First, to ensure that we could accurately estimate the time spent on each task, we filtered out participants who did not complete the tasks sequentially. This is because, for participants who completed tasks in a different order, it is unclear how much time they spent on each specific task. Sequential completion means the participant worked on the tasks in the order they were presented in the contest, allowing us to infer more accurate task-level time estimates.

Second, we applied standard methods to exclude outliers. Given that we assume the skills of participants are normally distributed, we used the common statistical technique of excluding data points that are more than two standard deviations from the mean.
   
\noindent \textbf{Debugging Energy Consumption (DEC)}: This metric quantifies the energy consumed during the time spent running code in the debugging phase. 

    First, we calculate the additional power consumed during debugging by calculating the runtime power (\ref{eq:runtime_power}) and excluding $p_\text{laptop}$ (which is already accounted for in the Coding Energy Consumption metric). The formula becomes:
    \[
    p_{\text{debug}} = p_{\text{runtime}} - p_{\text{laptop}}
    \]
    To estimate how long the code runs during debugging, we rely on existing studies. According to Stripe~\cite{stripe2018}, developers spend about 42\% of their total development time on debugging. Ko et al.~\cite{ko2006exploratory} found that, on average, only 10\% of the debugging time is spent actually running the code. Therefore, the time spent running code during debugging is calculated as:    
    \[
    t_{\text{debug}} = \text{Mean Time Spent} \times 0.42 \times 0.1
    \]

\noindent \textbf{Testing Energy Consumption (TEC)}: We estimated the energy consumption during testing based on the code's runtime. Similarly to CEC, we use Formula~\ref{eq:energy_formula} to calculate the testing energy consumption. However, as the value of power we use $p_{\text{runtime}}$, as during the testing process we need to actually run the code. 

To estimate the time the testing process takes, we need to know how much time one run of the tests takes and how many times the tests were run. For the former, we use the
run time data and calculate the average across all participants for each task. For the latter, we use the mean of the number of times a user submits their code. The values for both these metrics are provided by Codeforces and we accessed them using Codeforces API.

\subsection{Carbon Footprint for LLM-Assisted Approach}

To calculate the total energy consumption (TTEC) for the LLM-assisted approach we need to account for multiple components. The first one is the energy consumption associated with sending queries to an LLM ($E_{\text{query}}$). The second component is related to the fact that a human insight is part of our LLM-assisted approach (Step 4). Therefore, we need to calculate the energy consumption associated with this step by estimating the time it takes to produce the human insight ($t_{\text{insight}}$). Lastly, the process we have explained in the previous subsection does not always produce a solution that passes all test cases. Therefore, there is a need for the developer to understand the solution generated by the LLM and add the missing functionalities.
This makes for our third component to estimate which we need to know the time spent on adding missing functionalities ($t_{\text{add\_functionalities}}$). 

When accounting for all the mentioned components, our formula for total energy consumption becomes:
\[
E_{\text{total}} = E_{\text{query}} + (t_{\text{insight}} + t_{\text{add\_functionalities}}) \times p_{\text{laptop}}
\]

\noindent \textbf{Query Energy Consumption (QEC).} To estimate the overall query energy consumption we need to account for all queries we pass to the LLM and estimate the energy consumption of a single query. For the latter, we use the finding by Ludvigsen~\cite{ludvigsen2023chatgptenergy} that the energy consumption for each inference (query) is estimated to be 0.0022 kWh, based on the findings. In addition to the inference energy cost, we also account for the energy consumed during the training phase of the Large Language Model (LLM). To estimate the training energy cost per query, we divided the total training energy consumption (50 GWh) by the estimated number of queries that the LLM will handle over its lifecycle (5.68 Giga queries). Based on available data, we estimate the energy consumption per query for training to be 0.0088 kWh. Therefore, the total energy consumption per query, including both inference and training, is 0.011 (0.0022 + 0.0088) kwh.

It should be noted that these numbers are estimated using known hardware specifications and general data. No actual energy profiling tools or power meters were used to measure the precise energy consumption. We will discuss this limitation in more detail in the limitations section.

\noindent \textbf{Estimated Time Spent on Producing the Insight (ETHI)}: This metric accounts for the time a human spends understanding the task and providing insight to the LLM. According to a study by Minelli et al. \cite{minelli2015developersTime}, developers typically spend 38\% of their total time understanding a task.

Since the LLM already generates a partial solution, we adjust this time by considering the percentage of test cases passed by the LLM before receiving human insight. The idea is that the more the LLM succeeds initially, the less time the human needs to spend understanding and solving the remaining parts of the task.

The formula for estimating the time spent on producing the insight is:

    \[
    t_{\text{insight}} = t \times 0.38 \times (1 - TC_{\text{passed}})
    \]

    Where:
    \begin{itemize}
        \item $t$ is the total average time a developer would typically spend on the task,
        \item $0.38$ is the fraction of time developers typically spend on task understanding \cite{minelli2015developersTime},
        \item $TC_{\text{passed}}$ is the percentage of test cases passed before human insight, and $(1 - TC_{\text{passed}})$ represents the remaining portion of the task that needs human intervention.
    \end{itemize}

\noindent \textbf{Estimated Time to Add Missing Functionalities (ETAF)}: 
This metric estimates the time required to manually add any missing functionalities
when the LLM fails to fully complete the task, even after human insight is provided. The total time is a combination of the time spent on reading and extending the code, plus the time required to implement the remaining unsolved parts of the task. We exclude the time producing the insight as it means the developer already understands the task.

    The formula for estimating the time is:
    \[
    t_{\text{add}} = t_{\text{read\_extend}} + (1 - TC_{\text{after\_insight}}) \times t - t_{\text{insight}}
    \]

    Where:
    \begin{itemize}
        \item $t_{\text{read\_extend}}$ is the estimated time spent on reading and extending the code,
        \item $TC_{\text{after\_insight}}$ is the percentage of test cases passed after human insight has been given, $(1-TC_{\text{after\_insight}})$ represents the missing percentage of functionalities,
        \item $t$ is the total average time a developer spends on the task,
        \item $t_{\text{insight}}$ is the time spent on producing the human insight.
    \end{itemize}

To calculate $t_{\text{read\_extend}}$ we rely on the study by Ko et al. \cite{ko2006exploratory}, which reports that developers typically spend about 20\% of their debugging time reading code and another 20\% editing (extending) code. Thus, we estimate the time spent on reading and extending as a portion of the total debugging time. 

The formula for estimating the time is:

    \[
    t_{\text{read\_extend}} = t \times 0.42 \times (0.2 + 0.2)
    \]

    Where:
    \begin{itemize}
        \item $t$ is the total average time spent on the task,
        \item $0.42$ is the proportion of the total time spent on debugging ~\cite{stripe2018}
        \item $0.2$ is the fraction of debugging time spent on reading code,
        \item Another $0.2$ represents the fraction of debugging time spent on editing (extending) the code.
    \end{itemize}

\section{Results and Discussion} \label{sec:results}
In this section, we present and analyse the results obtained from our study. The analysis is structured around our two main research questions.
\subsection{RQ1: LLM-based vs. manual software development}
To answer RQ1, we need to compare the carbon footprint of the manual approach (based on the Codeforces contest data) to the carbon footprint of the LLM-based approach. Table \ref{table:manual1} presents the data for the manual process for the selected 12 tasks from 3 contests.
Table \ref{table:LLM}  presents the same data for the LLM-assisted approach. In each table we report data for three different contest that consist of 5, 4 and 3 tasks correspondingly. In the first column we list metrics associated with the calculation of the carbon footprint with their units indicated in the brackets.
As Table \ref{table:manual1} shows, the mean time spent (row "MTS") on each task varies on average between 444 to 2487 seconds, indicating the different level of effort the tasks take. 
\begin{table*}[t]
    \small
    \centering
    \caption {Manual Carbon Footprint Estimation for the selected contest rounds: each column represents a distinct task within a specific contest. For each task, the table reports Mean Time Spent (MTS), Coding Energy Consumption (CEC), Testing Energy Consumption (TEC), Debugging Energy Consumption (DEC), Total Energy Consumption (TTEC), and the resulting Carbon Footprint (CF). The measurement unit for each reported metric is indicated in brackets.}  
    \begin{tabularx}{\textwidth}{|X|X|X|X|X|X|X|X|X|X|X|X|X|X|X|l|l|}
    \hline
     &
     \multicolumn{5}{|c|}{\textbf{Contest 1983}} &
     \multicolumn{4}{|c|}{\textbf{Contest 1984}} &
     \multicolumn{3}{|c|}{\textbf{Contest 1994}}\\ \hline
         & A & B & C & D & E & A & B & C & D & A & B & C\\ \hline
        MTS
        (s) & 444 & 2216 & 2361 & 2018 & 2487 & 805 & 1437 & 1803 & 2146 & 748 & 1300 & 1783   \\ \hline
        
        CEC
        (kwh) & 5.03E-04
         & 2.51E-03
         & 2.67E-03
         & 2.28E-03
         & 2.82E-03
         & 9.11E-04
         & 1.63E-03
         & 2.04E-03
         & 2.43E-03
         & 8.47E-04
         & 1.47E-03
         & 2.02E-03\\ \hline
        TEC
        (kwh) & 3.37E-07 & 1.46E-06 & 3.49E-06 & 3.35E-06 & 9.64E-06 &6.82E-07 & 1.37E-06 & 2.59E-06 & 4.02E-06 & 8.41E-07 & 1.03E-06 & 2.26E-06  \\ \hline
        DEC
        (kwh) & 5.14E-05 & 2.57E-04 & 2.74E-04 & 2.34E-04 & 2.88E-04 & 9.32E-05 & 1.66E-04 & 2.09E-04 & 2.49E-04 & 8.66E-05 & 1.51E-04 & 2.07E-04  \\ \hline
        TTEC
        (kwh) &5.54E-04&2.77E-03&2.95E-03&2.52E-03&3.11E-03&1.01E-03&1.79E-03&2.25E-03&2.68E-03&9.34E-04&1.62E-03&2.23E-03
  \\ \hline
        CF (g) & 0.120&0.600&0.640&0.547&0.676&0.218&0.389&0.489&0.582&0.203&0.352&0.483

  \\ \hline
    \end{tabularx}
     \label{table:manual1}
\end{table*}

\begin{table*}[!t]
    \small
    \centering
    \caption {LLM-assisted Carbon Footprint Estimation for the selected contest rounds: each column represents a distinct task within a specific contest. For each task, the table reports  Number of Queries Before the Human insight (NQBH),  Number of Human Insight Queries (NHIQ), Percentage of the Test cases that Passed after the Human insight (TPAH),  Query Energy Consumption (QEC), Estimated Time Spent on Producing the Insight (ETHI), Estimated Time to Add Missing Functionalities (ETAF), Total Energy Consumption (TTEC) and the resulting Carbon Footprint (CF). The measurement unit for each reported metric is indicated in brackets.}
    \begin{tabularx}{\textwidth}{|X|X|X|X|X|X|X|X|X|X|X|X|X|X|X|l|l|}
    \hline
     &
     \multicolumn{5}{|c|}{\textbf{Contest 1983}} &
     \multicolumn{4}{|c|}{\textbf{Contest 1984}} &
     \multicolumn{3}{|c|}{\textbf{Contest 1994}}\\ \hline
     & A & B & C & D & E & A & B & C & D & A & B & C\\ \hline
        NQBH & 1 & 5 & 5 & 5 & 5 &4&5&5&5&1.67&5&5 \\ \hline
        NHIQ & 0.00 & 1.33 & 3.00 & 1.67 & 3.00 & 0.00	&2.00&3.00&3.00 & 0.00 &1.00&2.33 \\ \hline
        TPAH & 100\% & 100\% & 0\% & 69\% & 0\% &100\%	&100\%&73\%&69\%& 100\%&100\%&33\%
  \\ \hline
        QEC
        (Kwh) & 0.011 & 0.070 & 0.088 & 0.073 & 0.088 &0.044&0.077&0.088&0.088&0.018&0.066&0.081 \\ \hline
        ETHI
        (s) & 0 & 842 & 897 & 767 & 945 & 0&546&685&748&0&494&678  \\ \hline
        ETAF
        (s) & 0 & 0 & 1860 & 1459 & 1960 &0&0&109&282&
 0&0&817  \\ \hline
        TTEC
        (kwh) & 0.011 & 0.071 & 0.091 & 0.074 & 0.091 & 0.044&0.078&0.089&0.089& 0.018&0.067&0.082 \\ \hline
        CF (g) & 2.39&15.32&19.77&16.15&19.81& 9.55	&16.84&19.29&19.35&3.99&14.44&17.86
 \\ \hline
    \end{tabularx}
     \label{table:LLM}
\end{table*}

\begin{table*}[!t]
\small
\centering
    \caption{Ratio Difference between LLM and Manual Approach: each column represents a distinct task within a specific contest. The mean and standard deviation are reported across all tasks.
    }
    \label{tab:summaryRatio}
    \begin{tabularx}{\textwidth}{|X|X|X|X|X|X|X|X|X|X|X|X|X|X|X|l|l|}
    \hline
     &
     \multicolumn{5}{|c|}{\textbf{Contest 1983}} &
     \multicolumn{4}{|c|}{\textbf{Contest 1984}} &
     \multicolumn{3}{|c|}{\textbf{Contest 1994}}\\ \hline
     & A & B & C & D & E & A & B & C & D & A & B & C\\ \hline
   ratio & 19.92 & 25.53 & 30.89 & 29.52 & 29.30 & 43.81 & 43.29 & 39.45 & 33.25 & 19.66 & 41.02 & 36.98 \\ \hline
   \multicolumn{6}{|c|}{Mean:  32.72} &
   \multicolumn{7}{|c|}{Standard deviation: 8.41} \\ \hline
    \end{tabularx}
\end{table*}

In Table \ref{table:LLM} the row "NQBH" reports the number of queries sent to LLM before the human insight during the LLM-assisted software development. As the values in this row indicate, for only 3 tasks out of 12, the maximum number of 5 queries was not reached, indicating that LLMs were not able to solve the task (such that all provided test cases pass) without the support from human. The row "NHIQ" reports the number of human insight queries passed to LLM. We can see that for 4 tasks out of 12, this number also reached its maximum value of 3. The row "TPAH" reports the percentage of the test cases that passed after the human insight. This number is below 100\% for 6 tasks, i.e. for half of the tasks the LLM-assisted approach could not fully solve the task. 

When it comes to the overall carbon footprint, as it is clear from the results shown in Tables \ref{table:manual1} and \ref{table:LLM}, the LLM-assisted software development leads to a significantly higher carbon footprint compared to manual software development. Focusing on the total energy consumption and carbon footprint (last two rows of Tables  \ref{table:manual1} and \ref{table:LLM}), the LLM-based approach has at least 19.92 times and at most 43.81 times more carbon footprint (resp. energy consumption) than the manual approach (mean: 32.72, standard deviation: 8.41) The comparison of the two approaches is summarised in Table \ref{tab:summaryRatio}.

A closer look into the carbon footprint for the manual code generation shows that across 12 tasks the coding process accounts on average for 90.61\% of the carbon footprint, while testing and debugging account for 9.28\% and 0.11\% correspondingly. For LLM-assisted code generation, 98.68\% of the carbon footprint is due to querying the LLM. The remaining 0.77\% comes from generating human insights that support the LLM, while 0.56\% is due to developers adding missing functionalities. This suggests that optimizing LLM queries or incorporating more efficient models could significantly reduce the environmental impact of LLM-assisted development workflows.

\subsection{RQ2: The Impact of Task Complexity}

\begin{figure*}[!t]
    \centering
    \begin{subfigure}[b]{0.32\linewidth}
        \centering
        \includegraphics[width=\linewidth]{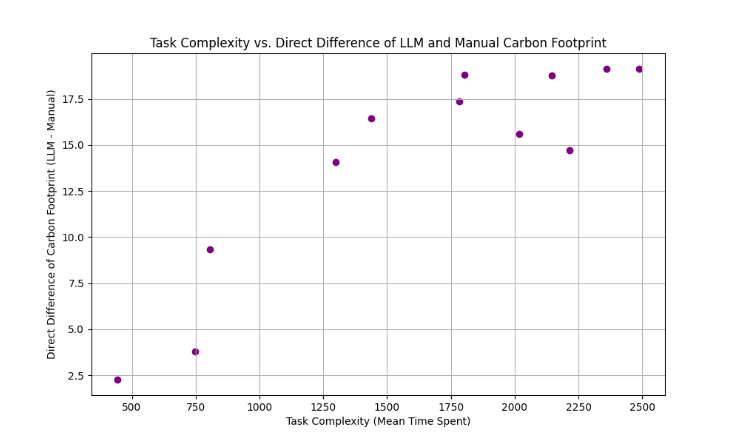}
        \caption{Task Complexity vs. Direct Difference of LLM and Manual Carbon Footprint}
        \label{fig:complexity_vs_difference}
    \end{subfigure}
    \hfill
    \begin{subfigure}[b]{0.32\linewidth}
        \centering
        \includegraphics[width=\linewidth]{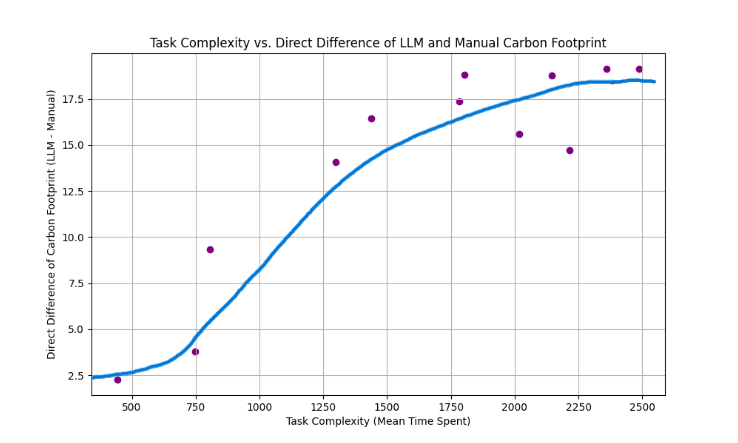}
        \caption{Task Complexity vs. Direct Difference of LLM and Manual Carbon Footprint with Expected pattern}
        \label{fig:expect}
    \end{subfigure}
    \hfill
    \begin{subfigure}[b]{0.32\linewidth}
        \centering
        \includegraphics[width=\linewidth]{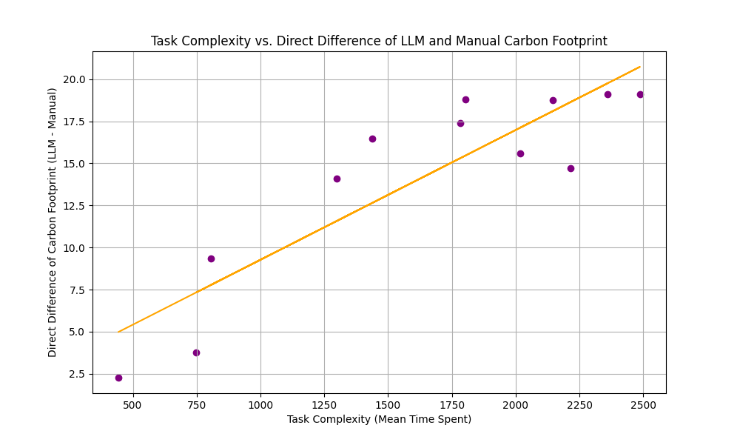}
        \caption{Task Complexity vs. Direct Difference of LLM and Manual Carbon Footprint with Best-Fit Line}
        \label{fig:complexity_vs_difference_best_fit}
    \end{subfigure}
    \caption{Comparison of Task Complexity and Carbon Footprint}
    \label{fig:three_graphs}
\end{figure*}

To address RQ2, i.e., to  investigate the correlation between task complexity and the difference in carbon footprints between the manual and LLM-assisted approaches, we first present a scatter plot \ref{fig:complexity_vs_difference} showing the relationship between task complexity (approximated by mean time spent) and the direct difference in carbon footprints between the manual and LLM-assisted approaches.

From our visual inspection of the scatter plot, it appears that as task complexity increases, the direct difference in the carbon footprint between LLM-assisted and manual approaches also increases.
The plot \ref{fig:expect} illustrates our expected pattern. We anticipate a relationship resembling a tanh function, where the direct difference in carbon footprint has lower and upper limits because of the number of maximum (8) and minimum (1) queries we set in our experiment.

To quantitatively assess the strength and significance of this relationship, we conducted two statistical tests: the Pearson correlation test and the Spearman rank correlation test with  significance level (\(\alpha\)) at 0.05. 
We tested the following hypotheses:

\begin{itemize}
    \item \textbf{Null Hypothesis (\(H_0\))}: There is no correlation between task complexity and the direct difference in carbon footprints between manual and LLM-assisted approaches.
    \item \textbf{Alternative Hypothesis (\(H_A\))}: There is a significant correlation between task complexity and the direct difference in carbon footprints between manual and LLM-assisted approaches.
\end{itemize}

The results of our statistical tests are as follows:

\begin{itemize}
    \item \textbf{Pearson Correlation Test:} The Pearson correlation coefficient was 0.890, with a p-value of 0.00011. Since the p-value is less than the significance level of 0.05, we reject the null hypothesis and conclude that there is a statistically significant positive linear correlation between task complexity and the direct difference in carbon footprints.
    
    \item \textbf{Spearman Rank Correlation Test:} The Spearman correlation coefficient was 0.840, with a p-value of 0.0006. This result also shows a significant monotonic relationship.
\end{itemize}

Based on these analyses, we conclude that there is a significant linear correlation between task complexity and the difference in carbon footprint of the two approaches. This finding emphasizes the potential drawbacks of using LLM for more complex tasks. 

\subsection{Best Practices for Green Coding with LLM }
In this section, we provide some recommendations that can inform future best practices for using LLM in green coding to achieve efficiency and minimal carbon footprint. 
To optimize the use of LLM in green software development, we recommend investigating the following approaches:
\begin{itemize}
\item  For complex tasks, decomposing them into smaller, manageable sub-tasks can mitigate the significant energy use gap (up to 30\% based on our observed data). This decomposition could be done manually or automatically, but the carbon footprint of the decomposition itself needs to be further investigated. 

\item We recommend assessing task complexity and adjusting the development process around it to minimise environmental impact. Depending on the task complexity we can determine the level of human involvement / autonomy in code generation. Other parameters such as the LLM-type and the resources used for training the LLM can further be used in fine-tuning the process once more data is made available about the public and open-source LLM, their training, and their carbon footprint.  
\end{itemize}

These practices can guide developers in making more informed decisions when using LLM in software development, balancing the benefits of LLM with the need to reduce environmental impact. 

\section{Threats to Validity}\label{sec:limitation}

We acknowledge several threats to the validity of our results and its generalisation. These limitations primarily arise from the limited data available to date and limited background research and methodologies.

\noindent \textbf{Scope of Analysis:} Our study focuses solely on the implementation, testing, and debugging phases of software development, excluding other stages such as planning and design. This limitation is due to the availability of data at the time of our research, and thus, our findings may not fully represent the entire software development life cycle.

\noindent \textbf{Assumptions about Hardware}: We assumed a specific machine configuration for the manual coding side of the study. Variations in the Thermal Design Power (TDP) of different machines could result in different energy consumption metrics. However, our conclusion on the statistical test should remain the same as our metrics in LLM also consider the manual effort.


\noindent 
\textbf{LLM and Prompting Limitations:} Our study is limited to the use of a single LLM, and as such, our findings may not be generalisable to other models with different architectures or training data. Additionally, we employed very simple zero-shot prompts—providing only the task description and asking the model to generate code without further guidance. While this reduces prompt-related bias, it may not reflect the full potential of LLMs under more sophisticated prompting strategies. Due to space constraints, the full set of prompts used is provided in our replication package \cite{replpackage}. 

\noindent \textbf{Limited Data Points}: While our analysis would benefit from more data points, including a broader range of tasks with different complexity, the limitations in the number of queries allowed with GPT-4 made it time-consuming to conduct the LLM experiments. 

\noindent \textbf{Estimation of Energy Consumption}: Many of the metrics, such as query energy consumption, are estimated using available hardware specifications of the LLM rather than precise energy profiling tools or power meters. Likewise, the total number of queries GPT-4 will handle over its lifetime is unknown but we estimated it using the lifetime of GPT-3.

\noindent \textbf{Competitive Programming Environment:} The tasks selected for our study are drawn from a competitive programming environment, which may not fully reflect real-world software development processes. As a result, the findings from this study might not directly translate to typical software development scenarios.

\section{Conclusions}\label{sec:conclusion}

In conclusion, we have shown that using an LLM-based approach to software engineering results in higher carbon emissions than a manual approach, with the gap increasing linearly as task complexity grows. 
Our work presents opportunities for a more in-depth analysis. For instance, we could delve into the reasons behind the difference in carbon emissions by conducting a detailed examination of LLM. This could involve studying the lifetime of an LLM or the training process, and investigating the impact of these variables. Such a comprehensive analysis could pave the way for making AI models, particularly LLM, more sustainable and energy efficient. 

\section{Acknowledgments}

This work has been supported by the ITEA grants GreenCode (project number 23016) and GENIUS (project number 23026).

\bibliographystyle{plain}
\bibliography{bare_conf}

\end{document}